\documentclass[twocolumn,floats,floatfix,aps,pra]{revtex4}
\usepackage{eurosym}
\usepackage{amsfonts}
\usepackage{amssymb}
\usepackage{amsmath}
\usepackage{graphicx}
\usepackage{bm}
\usepackage{color}
\usepackage{epsfig}
\usepackage{ifthen}
\usepackage{subfigure}
\usepackage{physics}
\usepackage[hidelinks, colorlinks=true,linkcolor=blue,citecolor=blue,filecolor=blue,urlcolor=blue]{hyperref}

\begin{document}

\title{Broadband composite nonreciprocal polarization wave plates and optical isolators}
\author{Hayk L. Gevorgyan}
\affiliation{Department of Physics, Sofia University, James Bourchier 5 blvd., 1164
Sofia, Bulgaria}
\author{Andon A. Rangelov}
\affiliation{Department of Physics, Sofia University, James Bourchier 5 blvd., 1164
Sofia, Bulgaria}
\author{Nikolay V. Vitanov}
\affiliation{Department of Physics, Sofia University, James Bourchier 5 blvd., 1164
Sofia, Bulgaria}

\begin{abstract}
We provide a technique for a broadband nonreciprocal wave retarder whose
quarter-wave plate phase retardation is the same in forward and backward
directions. The system is built using a number of sequential nonreciprocal wave plates. The proposed device can also be utilized to create a broadband optical diode, which consists of two achromatic quarter-wave plates, one reciprocal and the other non-reciprocal, that are sandwiched between two polarizers aligned in parallel.
\end{abstract}

\maketitle

\section{Introduction}

For decades, reciprocal and broadband (achromatic) polarization retarders
have been a topic of intense attention in optics \cite{Hecht,Wolf,Azzam,
Goldstein}. 
Traditionally, two or more conventional wave plates, of the same or different materials, are combined to make such retarders. West and Makas \cite{West}
reported achromatic combinations of plates with various birefringence
dispersions as one of the first known ideas. 
Destriau and Prouteau \cite{Destriau} presented achromatic retarders made out of wave plates of the same material but different thicknesses for two birefringent plates, while Pancharatnam offered three plates for half-wave \cite{Pancharatnam1} and quarter-wave \cite{Pancharatnam2} retarders. 
Harris and colleagues later presented achromatic quarter-wave plates with six \cite{Harris1} and ten identical quarter-wave plates \cite{Harris2}. 
The analogy between the polarization Jones vector and the quantum state vector has recently been used to suggest arbitrarily precise broadband polarization retarders \cite{Ardavan,Ivanov,Peters}.

All of the above achromatic wave plates are reciprocal, in the sense that their function is invariant upon time inversion. 
However, as recently shown by Al-Mahmoud et. al \cite{Mahmoud}, wave plates retarders can be non-reciprocal whose phase-shift retardation depends on the light propagation direction. 
For example, a retarder with retardation of $\pi /2$ in the forward direction (quarter-wave plate) and $\pi $ in the backward direction (half-wave plate) or other combination of retardance values can be
realized. 
The Al-Mahmoud et. al \cite{Mahmoud} non-reciprocal elements are based on magneto-optical phenomena like the Faraday effect. 
The axial (as opposed to polar) structure of the magnetic field and magnetization vectors in this case, as well as the associated invariance upon space inversion, are what cause the non-reciprocity. 
In Al-Mahmoud et. al \cite{Mahmoud} experiment, it was shown that a non-reciprocal Faraday rotator combined with a reciprocal rotator made of two half-wave plates sandwiched between crossed quarter-wave plates could be used to realize adjustable non-reciprocal wave retarders with retardation that differed in the forward and backward directions \cite{Mahmoud}.

In this paper, we theoretically propose novel broadband polarization
quarter-wave plates, which are also nonreciprocal, with the potential to be used in broadband optical isolators or/and circulators for
telecommunications, industrial, and laboratory research.

\section{Background}

The waveplate is a birefringent medium that modifies the polarization state by adding a phase shift of $\varphi$ between the two orthogonal
polarization components. The half-wave plate and quarter-wave plate
retarders are the most popular waveplates, with phase shifts of $\pi $ and $%
\pi /2$, respectively. The waveplate retarder's Jones matrix, whose axes are
aligned with the lab axes, takes the shape of a diagonal matrix,
\begin{equation}
J\left( \varphi \right) =\left[ 
\begin{array}{cc}
e^{i\varphi /2} & 0 \\ 
0 & e^{-i\varphi /2}%
\end{array}%
\right] ,
\end{equation}%
where $\varphi =2\pi L(n_{\mathnormal{s}}-n_{\mathnormal{f}})/\lambda $ is
the phase shift, $\lambda $ is the wavelength in vacuum, $n_{\mathnormal{f}}$
and $n_{\mathnormal{s}}$ are the refractive indices along the fast and slow
axes respectively, and $L$ is the thickness of the waveplate. When the
waveplate retarder's axes are rotated by an angle $\theta $ with regard to
the lab axes, the Jones matrix $J_{\theta }\left( \varphi \right) $ is given
by
\begin{widetext}
\begin{equation}
J_{\theta }\left( \varphi \right) =R\left( -\theta \right) J\left( \varphi
\right) R\left( \theta \right) =\left[ 
\begin{array}{cc}
e^{i\varphi /2}\cos ^{2}\left( \theta \right) +e^{-i\varphi /2}\sin
^{2}\left( \theta \right) & -i\sin \left( 2\theta \right) \sin \left(
\varphi /2\right) \\ 
-i\sin \left( 2\theta \right) \sin \left( \varphi /2\right) & e^{-i\varphi
/2}\cos ^{2}\left( \theta \right) +e^{i\varphi /2}\sin ^{2}\left( \theta
\right)%
\end{array}%
\right] ,
\end{equation}%
\end{widetext}
with rotation matrix $R\left( \theta \right) $ in the horizontal-vertical
(HV) basis given by
\begin{equation}
R\left( \theta \right) =\left[ 
\begin{array}{cc}
\cos \theta & -\sin \theta \\ 
\sin \theta & \cos \theta%
\end{array}%
\right].  \label{eq: rotator}
\end{equation}

Another way to realize a retarder is to use a polarization rotator at an
angle $\theta $ sandwiched in between two quarter-wave plates rotated by
angles $-\pi /4$ and $\pi /4$ with respect to the lab reference frame
correspondingly \cite{Messaadi}. The Jones matrix $J$ for such a sequence
can be given by the product of the Jones matrices of the quarter-wave plates
and the rotator: 
\begin{equation}
J=J_{-\pi /4}\left( \pi /2\right) R\left( \theta \right) J_{\pi /4}\left(
\pi /2\right) =\left[ 
\begin{array}{cc}
e^{i\theta } & 0 \\ 
0 & e^{-i\theta }%
\end{array}%
\right] =J_{0}\left( 2\theta \right) .  \label{eq: effective wave plate seq}
\end{equation}%
The last part of Eq.~(\ref{eq: effective wave plate seq})
demonstrates that the whole sequence can be considered an effective wave
plate with an effective retardation $\varphi =2\theta $. If one uses Faraday
rotator (nonreciprocal device) then the effective waveplate is also
nonreciprocal \cite{Mahmoud}. 
Even though the two quarter-wave plates can be achromatic --- an assumption we make from now on --- the effective wave plate is not broadband due to the strong wavelength dependence on the Verdet constant. 
Our objective in the present paper is to construct broadband nonreciprocal wave plates using a sequence of several nonreciprocal retarders, each with a specific phase shift and rotated by specific angles.

\section{Composite wave plate}

Now we will show three different sequences to construct nonreciprocal broadband quarter-wave plates. 

\begin{itemize}

\item The first approach is to combine two nonreciprocal quarter-wave plates and one nonreciprocal half-wave plate.
This composition is described by the Jones matrix
\begin{equation}
\mathcal{J}\left( \varepsilon \right) =J_{\alpha _{1}}\left( \pi
/2+\varepsilon /2\right) J_{\alpha _{2}}\left( \pi +\varepsilon \right)
J_{\alpha _{3}}\left( \pi /2+\varepsilon /2\right) .  \label{simpel1}
\end{equation}%

\item
The second approach is to combine two nonreciprocal half-wave plates and one
nonreciprocal quarter-wave plate, characterized by the Jones matrix 
\begin{equation}
\mathcal{J}\left( \varepsilon \right) =J_{\alpha _{1}}\left( \pi
+\varepsilon \right) J_{\alpha _{2}}\left( \pi +\varepsilon \right)
J_{\alpha _{3}}\left( \pi /2+\varepsilon /2\right) .  \label{simpel2}
\end{equation}

\item The third approach is to have multiple nonreciprocal wave plates in the sequence, e.g., combining four nonreciprocal half-wave plates and one nonreciprocal quarter-wave plate.
The Jones matrix of this structure reads
\begin{align}
\mathcal{J}\left( \varepsilon \right) &= 
J_{\alpha _{1}}\left( \pi +\varepsilon \right) 
J_{\alpha _{2}}\left( \pi +\varepsilon \right)
J_{\alpha _{3}}\left( \pi +\varepsilon \right) \notag\\ 
&\times J_{\alpha _{4}}\left( \pi +\varepsilon \right) 
J_{\alpha _{5}}\left( \pi /2+\varepsilon /2\right) .
\label{simpel3}
\end{align}%
Here $\varepsilon $ and $\varepsilon /2$\ represent the systematic deviations from the nominal retardation of the half- and quarter-wave plates respectively.

\end{itemize}

We note that a combination of only reciprocal elements leads to reciprocal sequence, but the combination of reciprocal and nonreciprocal elements may lead to reciprocal or nonreciprocal sequences. 
For the above odd number of sequences (\ref{simpel1})-(\ref{simpel3}) one can easily check that they are nonreciprocal.

The composite retarder's efficiency is evaluated in terms of the fidelity $\mathfrak{F}$  \cite{Ardavan},
\begin{equation}
\mathfrak{F}\left( \varepsilon \right) =\frac{1}{2}\left\vert \Tr\left(
J_{0}^{-1}\mathcal{J}\left( \varepsilon \right) \right) \right\vert ,
\label{fidelity}
\end{equation}
where $\mathcal{J}\left( \varepsilon \right) $ is the achieved and $J_{0}$ is the target Jones matrix. $\mathfrak{F}=1$ if the two operators $\mathcal{J}$ and $J_{0}$ are identical, but the fidelity reduces if the two matrices differ. 
In order to produce broadband nonreciprocal quarter-wave plate we determine the rotation angles of each wave plate in Eqs.~(\ref{simpel1}), (\ref{simpel2}) or (\ref{simpel3}) by using the Monte Carlo method. We generate 10$^{4}$ sets of random angles $\alpha _{1},\alpha _{2},\alpha_{3},\alpha _{4}$ and $\alpha _{5}$. 
We select solutions, which deliver the biggest overall fidelity $\mathfrak{F}\left( \varepsilon \right) $ in the interval of $\varepsilon \in \left[ -\pi ,\pi \right] $ and also, ensure a flat top. 
The angles are presented in table \ref{table1}. 
It is important to note that the parameters given in table \ref{table1} are not the only possible. 
Obviously, for the wavelength at which the wave plates serve as half or quarter-wave plates, respectively, we have $\varepsilon =0$ and $\mathfrak{F}\left( 0\right) =1$.

\begin{table}[tbh]
\caption{Calculated angles of rotation (in radians) for the three sequences of Eqs.~(\protect\ref{simpel1}), (\protect\ref{simpel2}), and (\protect\ref{simpel3}).}
\label{table1}%
\begin{tabular}{|c|c|}
\hline
sequences & angles $(\alpha _{1};\alpha _{2};\ldots;\alpha _{N})$ \\ \hline
(\ref{simpel1}) & (3.3; 1.21; 3.1) \\ \hline
(\ref{simpel2}) & (3.6; 1.65; 3.9) \\ \hline
(\ref{simpel3}) & (1.61;6.48;6.47;1.62;0.78) \\ \hline
\end{tabular}%
\end{table}

\section{Broadband optical isolator}

\begin{figure}[htb]
\centerline{\includegraphics[width=1\columnwidth]{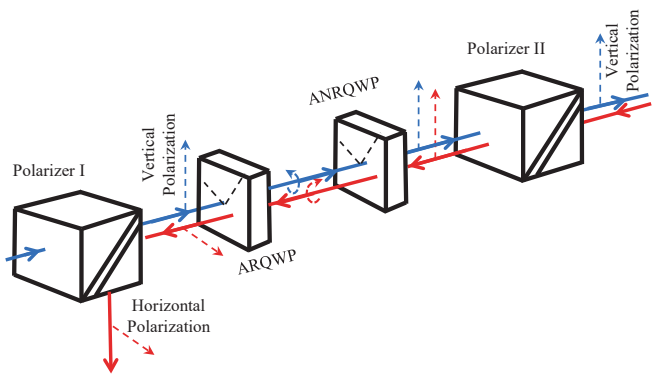}}
\caption{Scheme of the broadband optical isolator. ARQWP stands for
the achromatic reciprocal quarter-wave plate, while ANRQWP stands for the achromatic
nonreciprocal quarter-wave plate.}
\label{fig1}
\end{figure}

Another interesting case is when the sequence serves as a broadband null
retarder in one direction and a broadband half-wave plate in the other
direction, which can be archived if we combine our nonreciprocal broadband
quarter-wave plate with a commercially available broadband but reciprocal quarter-wave
plate. 
In this case, one can build a broadband optical isolator as shown and explained in Fig. \ref{fig1}.

The working principle of the proposed optical isolator  is the following.
Any light beam entering through the polarizer
I will exit vertically polarized (blue array), after passing through the
achromatic reciprocal quarter-wave plate (ARQWP) the light will be
circularly polarized, then passing through the achromatic nonreciprocal
quarter-wave plate (ANRQWP) it will be again vertically polarized, thus all
light will pass polarizer II. On the way back, if the light re-enters the
polarizer II in the backward direction (red array), due to the combined effect
of the ANRQWP and ARQWP, the polarization is rotated in such a way (90 degrees) that the
whole wave is blocked by the polarizer I, so that no light can exit from
right to left.

\section{Numerical calculations}

\begin{figure}[htb]
\includegraphics[width=1\columnwidth]{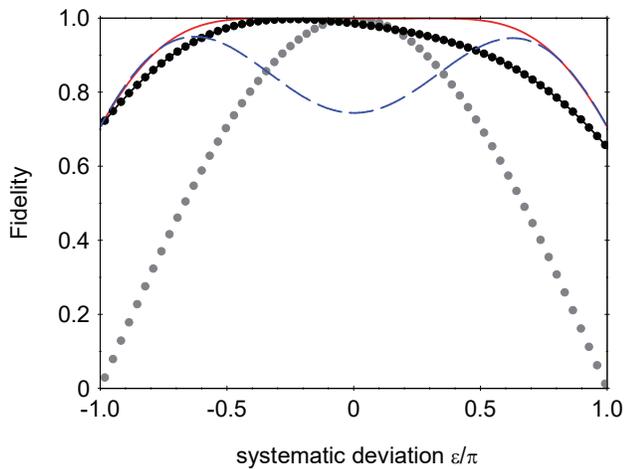}
\caption{Fidelity versus systematic deviation for the composite
waveplates designed by using three configurations: Eq.~\eqref{simpel1} is depicted by the blue dashed line, Eq.~\eqref{simpel2} by the black dotted line, and Eq.~\eqref{simpel3} by the red solid line. 
The gray dotted line is for a quarter-wave plate with a single Faraday rotator for easy reference. }
\label{fidelity figure}
\end{figure}

We explained the basic concept of creating broadband nonreciprocal polarization quarter-wave plates and broadband optical isolators in the previous sections. 
Now, we present numerical simulations to test the effectiveness of the design we have discussed above.

In Figure \ref{fidelity figure} we show the calculation for the fidelity $\mathfrak{F}$ profiles using the three configurations (\ref{simpel1}), (\ref{simpel2}) and (\ref{simpel3}) with rotation angles taken from table \ref{table1}. 
Obviously, the configuration (\ref{simpel3}) outperforms the other
configurations and this is expected because configuration (\ref{simpel3}) has five retarders in the series compared to three retarders in case of (\ref{simpel1}) and (\ref{simpel2}). 
In theory, the fidelity profiles can be made arbitrarily flat by increasing the number of retarders in the series. 
In practice, it is not clear whether such many-retarders sequences will be useful, due to the many optical elements in the series (Faraday rotators and quarter-wave plates), therefore we limit our investigation to five nonreciprocal quarter-wave plates (five Faraday rotators and ten
quarter-wave plates altogether).

For broadband optical isolator simulations in this paper, we use terbium gallium garnet crystal (TGG) as it is one of the most common crystals for Faraday rotators. 
We fix the applied magnetic field to 1 T, the length of the crystal is considered to be 1 cm for the half-wave plates and 0.5 cm for the quarter-wave plates. 
Up until now, there has been a lot of research done on the dispersion of the TGG Verdet constant $\nu $ \cite{Bozinis1978, Jannin1998, Yoshida}. 
It was demonstrated that the following formula can adequately represent the wavelength dependence of this crystal,
\begin{equation}
\nu (\lambda )=\frac{K}{\lambda _{0}^{2}-\lambda ^{2}}\,,  \label{dispersion}
\end{equation}%
where $K=4{.}45\cdot 10^{7}\,\frac{\text{rad $\cdot $ nm}^{2}}{\text{T $\cdot $ m}}$ and $\lambda _{0}=258{.}2$~nm is the effective transition wavelength. TGG has optimal material properties for the Faraday rotator in the range of 400 -- 1100 nm, excluding 470 -- 500 nm (the absorption window). For most materials, the Verdet constant decreases (in absolute value) with increasing wavelength: for TGG it is equal to $134\,\frac{\text{rad}}{\text{T $\cdot $ m}}$ at $632$~nm and $40\,\frac{\text{rad}}{\text{T $\cdot $ m}}$ at $1064$~nm. The operating wavelength range of the Faraday isolator is constrained as a result of this.

The performance of the optical isolators is quantified by its transmission $T_{f}$ (a portion of the input light's intensity that passes through the isolator), back-transmission $T_{b}$ (a portion of the back-transmission light's intensity that passes through the isolator in the opposite direction), and isolation $D$. The light intensity measured after passing the optical diode in both the forward ($I_{forw}$) and backward ($I_{back}$) directions determine these numbers, respectively,
\begin{subequations}
\label{intensities}
\begin{eqnarray}
T_{f}=I_{forw}/I_{0} &=&\left\lvert \mathbf{P}_{V}\mathbf{J}_{f}\mathbf{P}%
_{V} \ket{in}\right\vert ^{2},  \label{forward intensity} \\
T_{b}=I_{back}/I_{0} &=&\left\vert \mathbf{P}_{V}\mathbf{J}_{b}\mathbf{P}%
_{V} \ket{in} \right\vert ^{2},  \label{backward intensity}
\end{eqnarray}%
where $\mathbf{P}_{V}$ stand for vertical polarizers, $|in\rangle $ is the
Jones vector for the light entering the isolator, and $\mathbf{J}_{f}$ and $%
\mathbf{J}_{b}$ are Jones matrices for forward and backward wave plates
respectively. $I_{0}$ has the meaning of the intensity of light at the
beginning of the Faraday isolator.
The isolation is then determined with the formula \cite{Kim2006,Adams2012}
\end{subequations}
\begin{equation}
D=-10\log \left[ \frac{T_{b}}{T_{f}}\right] \,.  \label{isolation}
\end{equation}

\begin{figure}[htb]
\centerline{\includegraphics[width=1\columnwidth]{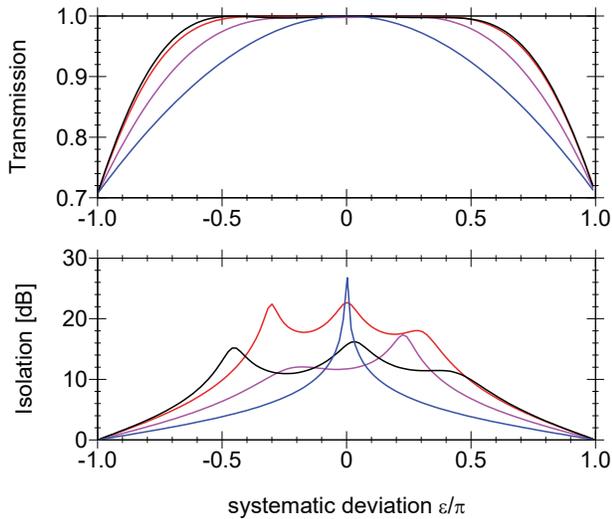}} 
\caption{Transmission and isolation properties of the optical isolators with different numbers of wave plates in the series, compared to the isolator based on a single rotator (blue line), vs the systematic deviation $\protect\varepsilon $. 
The other three curves refer to the sequences of Eqs.~(\protect\ref{simpel1}) depicted by a purple line, (\protect\ref{simpel2}) by a red line, and (\protect\ref{simpel3}) by a black line.}
\label{Fig3}
\end{figure}

\begin{figure}[htb]
\centerline{\includegraphics[width=1\columnwidth]{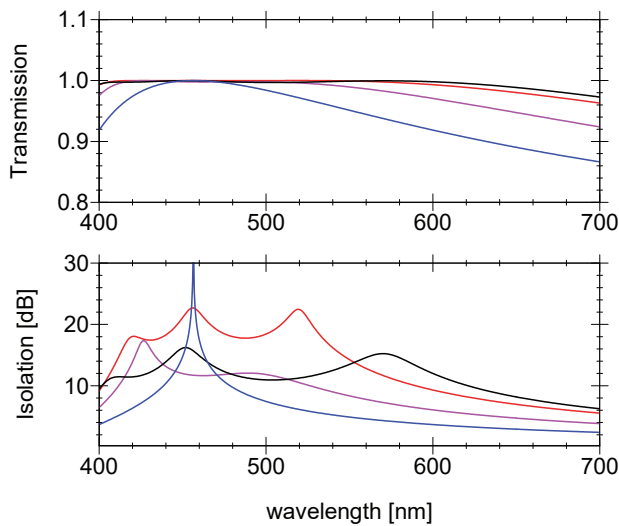}}
\caption{The same as Fig.~\protect\ref{Fig3} but instead of systematic deviation $\protect\varepsilon $ we use the wavelength parameter. 
}
\label{Fig4}
\end{figure}

The transmission and isolation profiles for the three configurations (\ref{simpel1}), (\ref{simpel2}) and (\ref{simpel3}) are shown in Figs.~\ref{Fig3} and \ref{Fig4}. 
One can notice that for all these composite isolators both the transmission and isolation are far more efficient than that of isolators using a single rotator (blue curve). 
Figure~\ref{Fig3} shows the performance of the optical isolators under study with respect to the systematic deviation $\varepsilon $ of the Faraday rotators, whereas in Fig.~\ref{Fig4} the analogous dependence on wavelength is presented. 
The long tail asymmetry seen in Fig.~\ref{Fig4} stems from the fact that the Faraday rotation angle depends non-linearly on the wavelength (as seen in Eq.~(\ref{dispersion})).
The isolation above 10 dB over a region of 200 nm can be seen from Fig.~\ref{Fig4}, and it is a much broader spectral range compared to the case of using just a single Faraday rotator (about 20 nm on the level of 10 dB).

We emphasize that the transmission and isolation curves were calculated under the assumption of no losses in realistic realizations.
Insertion losses and reflections from the surfaces of the optical elements are the primary causes of the reduced transmission. 
Losses resulting from optical element propagation would not be as significant.

\section{Conclusion}

We have presented a novel way to construct broadband nonreciprocal polarization quarter-wave plates. 
The concept is based on combination of several nonreciprocal waveplates with the optical axis of
each rotated by appropriate angles. 
In addition, the proposed broadband nonreciprocal polarization quarter-wave plate can be used in combination with a broadband reciprocal polarization quarter-wave plate to build a broadband optical isolator. 
The isolation bandwidth (isolation of more than 10 dB) is almost 200 nm while the transmission bandwidth is beyond 200 nm. 
This isolator has the benefit of being resistant to changes in temperature, crystal length, and magnetic field. 
With the available optical components, an experimental implementation should be feasible. 

\section*{Acknowledgments}

This research is partially supported by Bulgarian national plan for recovery
and resilience, contract BG-RRP-2.004-0008-C01 SUMMIT: Sofia University
Marking Momentum for Innovation and Technological Transfer, project number
3.1.4. and the University of Sofia Grant 80-10-26/20.04.2023.


\end{document}